\documentstyle[12pt]{article}

\advance\textheight by \topskip
\begin{document}
\tolerance=100000
\thispagestyle{empty}
\setcounter{page}{0}

\newcommand{\be}{\begin{equation}}
\newcommand{\ee}{\end{equation}}
\newcommand{\br}{\begin{eqnarray}}
\newcommand{\er}{\end{eqnarray}}
\newcommand{\ba}{\begin{array}}
\newcommand{\ea}{\end{array}}
\newcommand{\bi}{\begin{itemize}}
\newcommand{\ei}{\end{itemize}}
\newcommand{\bn}{\begin{enumerate}}
\newcommand{\en}{\end{enumerate}}
\newcommand{\bc}{\begin{center}}
\newcommand{\ec}{\end{center}}
\newcommand{\ul}{\underline}
\newcommand{\ol}{\overline}
\newcommand{\eebbww}{$e^+e^-\rightarrow b\bar b W^+W^-$}
\newcommand{\ar}{\rightarrow}
\newcommand{\sm}{${\cal {SM}}$}
\newcommand{\mssm}{${\cal {MSSM}}$}
\newcommand{\Dir}{\kern -6.4pt\Big{/}}
\newcommand{\Dirin}{\kern -10.4pt\Big{/}\kern 4.4pt}
\newcommand{\DDir}{\kern -7.6pt\Big{/}}
\newcommand{\DGir}{\kern -6.0pt\Big{/}}

\def\Ord{\buildrel{\scriptscriptstyle <}\over{\scriptscriptstyle\sim}}
\def\OOrd{\buildrel{\scriptscriptstyle >}\over{\scriptscriptstyle\sim}}
\def\pl #1 #2 #3 {{\it Phys.~Lett.} {\bf#1} (#2) #3}
\def\np #1 #2 #3 {{\it Nucl.~Phys.} {\bf#1} (#2) #3}
\def\zp #1 #2 #3 {{\it Z.~Phys.} {\bf#1} (#2) #3}
\def\pr #1 #2 #3 {{\it Phys.~Rev.} {\bf#1} (#2) #3}
\def\prep #1 #2 #3 {{\it Phys.~Rep.} {\bf#1} (#2) #3}
\def\prl #1 #2 #3 {{\it Phys.~Rev.~Lett.} {\bf#1} (#2) #3}
\def\mpl #1 #2 #3 {{\it Mod.~Phys.~Lett.} {\bf#1} (#2) #3}
\def\rmp #1 #2 #3 {{\it Rev.~Mod.~Phys.} {\bf#1} (#2) #3}
\def\ijmp #1 #2 #3 {{\it Int.~J.~Mod.~Phys.} {\bf#1} (#2) #3}
\def\sjnp #1 #2 #3 {{\it Sov.~J.~Nucl.~Phys.} {\bf#1} (#2) #3}
\def\xx #1 #2 #3 {{\bf#1}, (#2) #3}
\def\preprint{{\it preprint}}

\begin{flushright}
{\large DFTT 60/94E}\\
{\large DTP/94/104E}\\
{\rm October 1995\hspace*{.5 truecm}}\\
\end{flushright}

\vspace*{\fill}

\begin{center}
{\Large \bf
Contributions of below--threshold decays to ${\cal {MSSM}}$
Higgs branching ratios: Erratum\footnote{Work supported in part by Ministero
dell' Universit\`a e della Ricerca Scientifica.\\[4. mm]
E-mail: Moretti@hep.phy.cam.ac.uk; W.J.Stirling@durham.ac.uk.}}\\[2.cm]
{\large
S.~Moretti$^{a,b}$ and W.~J.~Stirling$^{c}$}\\[0.5 cm]
{\it a) Dipartimento di Fisica Teorica, Universit\`a di Torino,}\\
{\it and I.N.F.N., Sezione di Torino,}\\
{\it Via Pietro Giuria 1, 10125 Torino, Italy.}\\[0.5cm]
{\it b) Cavendish Laboratory, University of Cambridge,}\\
{\it Madingley Road, Cambridge CB3 0HE, United Kingdom.}\\[0.5cm]
{\it c) Departments of Physics and
Mathematical Sciences, University of Durham,}\\
{\it South Road, Durham DH1 3LE, United Kingdom.}\\[0.75cm]
\end{center}

\vspace*{\fill}

\begin{abstract}
{\normalsize\noindent In Ref.~\cite{paper} we calculated all the
experimentally relevant branching ratios of the Higgs bosons of the
Minimal Supersymmetric Standard Model, paying
particular attention to the contributions from below--threshold
decays. Unfortunately, an error in one of the subroutines of the {\tt FORTRAN}
code we used was affecting the computation of the off--shell partial widths
of the decays $A\ar Z^{0*}h^*$ and $H^\pm\ar W^{\pm *}h^*$.
This has now been fixed, and the corrected plots are presented here.}
\end{abstract}

\vspace*{\fill}
\newpage

\section*{Corrected results for the $A\ar Z^{0*}h^*$ and
$H^\pm\ar W^{\pm *}h^*$ branching ratios
at small $\tan\beta$}

The error in the program was affecting the two
decay channels $A\ar Z^{0*}h^*$ and
$H^\pm\ar W^{\pm *}h^*$ below the real particle
 thresholds at $M_{Z^0}+M_{h}$ and
$M_{W^\pm}+M_{h}$ respectively. The corresponding rates for
the on--shell decays were correct.
The overall effect was to underestimate  the off--shell partial widths
(and consequently  the branching ratios): this was
substantial at small values of
$\tan\beta$, but negligible at large values  since in this
latter case the two channels are heavily suppressed.
Figs.~3 and 5 (which replace the corresponding figures in
Ref.~\cite{paper}) show the new results. The comments
in the text of Ref.~\cite{paper} remain unchanged. We note that
the corrected rates have phenomenological relevance for the case
$H^\pm\ar W^{\pm *}h^*$, whereas for $A\ar Z^{0*}h^*$ the impact is
largely reduced.

Our results agree now with the
rates given in Ref.~\cite{alter}, within  computational errors
and taking into account  different choices of the
parameters, scales, etc.
The {\tt FORTRAN} code  used in our analysis is available on request
from the authors.

\section*{Acknowledgements}

\noindent We thank A.~Djouadi, J.~Kalinowski and P.M.~Zerwas for pointing
out the error.


\section*{Figure captions}

\begin{itemize}
\item[{[3]}] Branching ratios for the pseudoscalar
\mssm\ Higgs bosons $A$ as a function of $M_A$, for $\tan\beta = 1.5$
and 30. Other parameter values are given in the text of Ref.~\cite{paper}.
\item[{[5]}] Branching ratios for the charged
\mssm\ Higgs bosons $H^\pm$ as a function of $M_{H^\pm}$, for $\tan\beta = 1.5$
and 30. Other parameter values are given in the text of Ref.~\cite{paper}.
\end{itemize}
\vfill
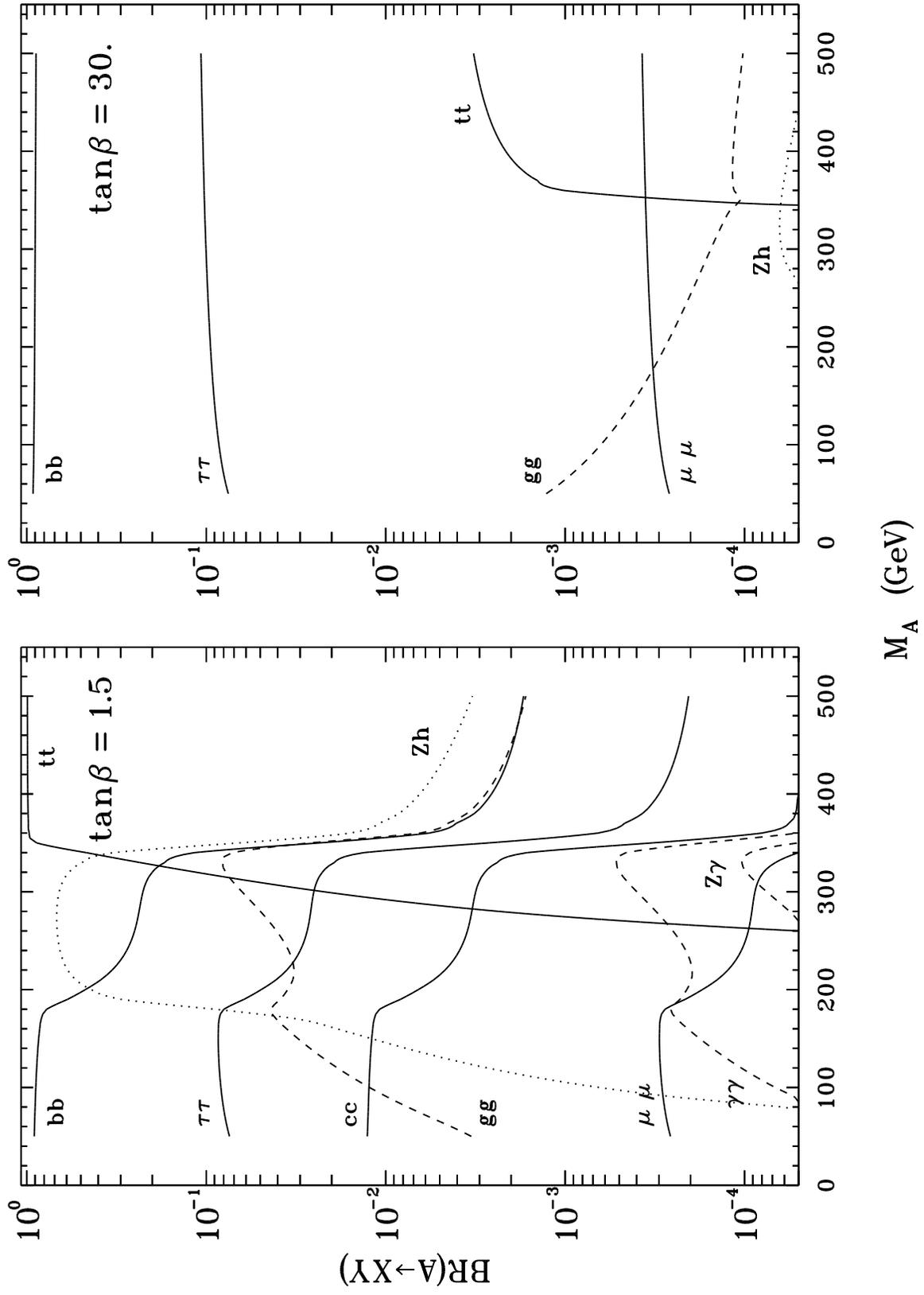
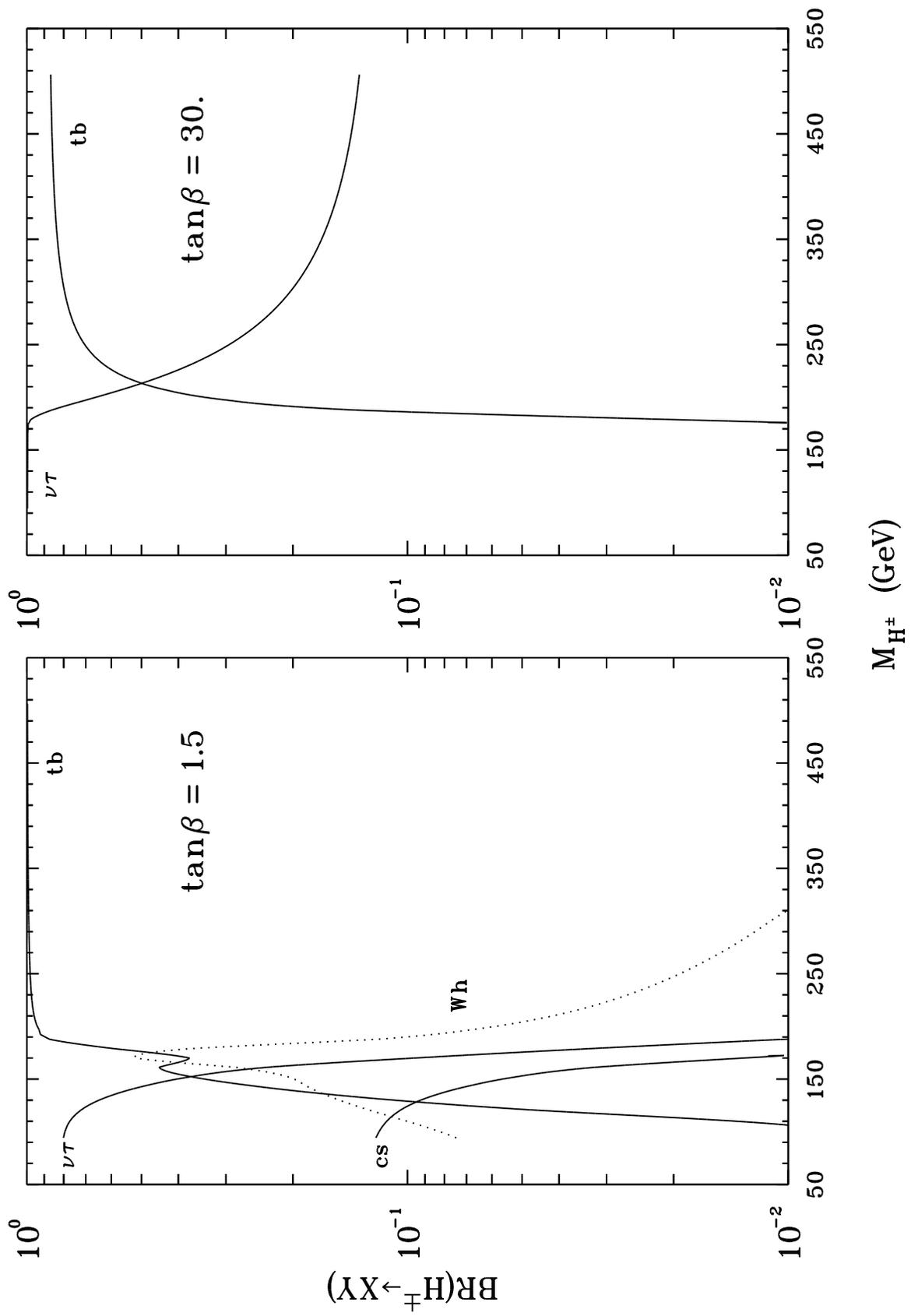
\end{document}